# Detection Sensitivity Limit of Hundreds of Atoms with X-Ray Fluorescence Microscopy

M. G. Masteghin[1], T. Gervais[1], S. K. Clowes[1], D. C. Cox[1], V. Zelyk[1],
A. Pattammattel[2], Y. S. Chu[2],
N. Kolev[3], T. Z. Stock[3], N. Curson[3],
P. G. Evans[4],
M. Stuckelberger[5],
B. N. Murdin[1]*

[1] *Advanced Technology Institute, University of Surrey, Guildford GU2 7XH, UK*
[2] *National Synchrotron Light Source II, Brookhaven National Laboratory, Upton, NY 11973, USA*
[3] *London Centre for Nanotechnology, University College London, London, WC1H 0AH, UK*
[4] *Department of Materials Science and Engineering, University of Wisconsin-Madison, Madison, Wisconsin 53706, United States*
[5] *Center for X-ray and Nano Science CXNS, Deutsches Elektronen-Synchrotron DESY, Notkestraße 85, 22607 Hamburg, Germany*
* Corresponding author's e-mail: b.murdin@surrey.ac.uk

**ABSTRACT.** We report X-ray fluorescence (XRF) imaging of nanoscale inclusions of impurities for quantum technology. A very bright diffraction-limited focus of the X-ray beam produces very high sensitivity and resolution. We investigated gallium (Ga) dopants in silicon (Si) produced by a focused ion beam (FIB). These dopants might provide 3/2-spin qubits or p-type electrical contacts and quantum dots. We find that the ion beam spot is somewhat larger than expected, and the technique provides a useful calibration for the resolution of FIBs. Enticingly, we demonstrate that with a single shot detection of 1 second integration time, the sensitivity of the XRF would be sufficient to find amongst background a single isolated inclusion of unknown location comprising only 3000 Ga impurities (a mass of just 350 zg) without any need for specialized nm-thickness lamellae, and down from >10$^5$ atoms in previous reports of similar work. With increased integration we were able to detect 650 impurities. The results show that planned facility upgrades might achieve single atom sensitivity with a generally applicable, non-destructive technique in the near future.

## I. INTRODUCTION

Impurities in solids are important for the (opto-)electronic properties they give, and canonical semiconductor dopants are a prime example. Recently, there has been a drive towards controlled incorporation of single impurities for quantum technologies, and in the case of silicon dopants these impurities carry a spin for use in quantum sensing or computation or memory[1]. In other materials, single photon emission in quantum communications is provided by impurities and defect complexes[2]. Ion implantation is frequently used in such technologies and a variety of strategies is being pursued to ensure single impurities are incorporated with high positioning precision with focussed ion beams[3-6]. In none of these cases has a single ion been detected post-implant and been verifiably shown to be produced by deterministic incorporation. For example, multiple ions have been deliberately implanted at each site because of the multiple possible host sites[3], or where single defects have been detected, their occurrence is random[2]. Even when quantum devices are produced based on a precise number of impurities, the location of those impurities and the number of unintentionally implanted impurities not involved in the device operation is not described[7]. One extremely precise, deterministic lithography technique based on scanning probes has been established[8], but only for two species of impurity (phosphorus and arsenic) in one host (silicon), and ion implantation is more rapid and flexible. Imaging of single impurities for post-implant validation with chemical specificity is challenging and has so far been established only for specialised X-ray techniques involving very thin lamellae[9-11] and surfaces[12]. X-ray Fluorescence (XRF) may provide a crucial tool, since it allows chemically selective imaging on a nm scale and does not require special (destructive) sample preparation. Regarding implantation of materials for quantum technology, XRF does not rely on

.

activation of the impurity and incorporation into an optically addressable defect, which can be far below the surface. It has already been used to detect deterministically incorporated impurities[13] and biomedically relevant trace elements[14], although only with micron scale resolution, and a sensitivity sufficient to detect only $10^5$ or more atoms. Single atom imaging has been achieved with other techniques, but none have the generality of application of XRF, which would be very attractive. For example, single defect fluorescence at visible wavelengths is common[2] (as is single molecule fluorescence[15]), but this is restricted to specific defects/molecules with high luminescence yield and has resolution limited by the much longer wavelengths used.

In this work, we investigate the sensitivity limits of a non-destructive and generally applicable XRF microscope. We chose to investigate Ga implants because they are very easy to produce with controlled density in high resolution patterns, and we used silicon as the host because it can be obtained with very high chemical purity. Ga in silicon also has a conveniently large mass difference meaning that there is good contrast between the XRF photon energies of the impurity and host. This combination is not particularly favoured as a quantum technology, but it could provide a 3/2-spin qubit and single Ga placement has been attempted to that end[16], and it allows a convenient demonstration of the principles.

In an XRF experiment, fluorescence photons are counted and binned according to their energy, which allows identification of the source elements in the beam. The identification is confounded by background consisting primarily of signals from other elements with nearby XRF energies, which may come from the sample or the environment (such as gas in the sample chamber or the X-ray optics etc). The ability to identify a pixel containing a specific chemical species requires a choice of a count threshold, $C$, such that pixels with count $c > C$ are identified as containing impurities and those with $c < C$ produce emission only from background. The choice of $C$ should be such that there is a low probability of both false positives, $p(\text{false} + \text{ve}) = p(c < C|0)p(0)$, and false negatives, $p(\text{false} - \text{ve}) = \sum_N p(c < C|N)p(N)$ where $N$ is the number of impurities. If $p(c|N)$ is Poissonian with mean $\langle c(N) \rangle$, we can calculate the conditional probabilities for false positives and negatives for a given threshold: $p(c > C|0) = 1 - \Gamma(C, \langle c(0) \rangle)/\Gamma(C)$ and $p(c < C|N) = \Gamma(1 + C, \langle c(N) \rangle)/\Gamma(1 + C)$, in which $\Gamma(z)$ is the Euler Gamma function and $\Gamma(a, z)$ is the incomplete Gamma function. A full discussion of the optimization of $C$ is application specific because the prior probabilities of positives, $p(N)$, and negatives, $p(0)$, i.e., the proportion of pixels containing impurities (and whether or not they are in a pattern) is sample specific. For a simplified discussion, let us suppose that we require that the probability distributions for $p(c|N)$ and $p(c|0)$ should have small overlap, which we can specify from:

$$z(N) = \frac{\langle c(N) \rangle - \langle c(0) \rangle}{\sqrt{\sigma_N^2 + \sigma_0^2}}$$

Pixels containing impurities can be identified correctly if $z$ is large. For example, requiring that the distributions are $2\sigma$ apart would be obtained with $z=2$, which gives a one-tailed test probability of 2.3% that we might use as a proxy for the false detection error rate. The objective of this study is to provide $p(c|N)$ and $\langle c(N) \rangle$.

## II. EXPERIMENTAL PROCEDURE

### II.1. Samples

The sample was prepared from a silicon-on-insulator (SOI) substrate (SOITEC). The top silicon (device) layer had a thickness of 450 nm and a buried oxide (box) layer of 1000 nm. The device layer is boron doped with a resistivity of $10^4$ ohm cm. Initially, for ease of navigation to regions of interest, chromium (Cr) markers were created via positive electron beam lithography (EBL). We used baked poly(methyl methacrylate), PMMA A6, exposed to small electron doses that following development has delimited regions where a 30 nm Cr layer was deposited by e-beam evaporation. Subsequently, singly charged gallium ions (Ga$^+$) with an energy of 30 keV were implanted using a dual beam Focused Ion Beam microscope and Scanning Electron Microscope (FIB-SEM, FEI Nova Nanolab). The ion current was measured before and after the implantation with a Faraday cup and determined to be $i$=2.4 pA on both readings and, therefore, was assumed constant throughout the implantation and to have an uncertainty less than 4%. Simulations (SUSPRE, UKNIBC)[17] show that the implantation of Ga at this energy produces a straggle with rms radius of 9 nm, and the rms spot size of the FIB was nominally 15nm. The implanted areas were produced by implanting spots with a pitch of $p$=15.64 nm, i.e. similar to the nominal radial distribution of Ga ions from each spot. The average 2D dose, n, in impurity ions per unit area was determined with the dwell time for each spot, $t$, from $n = it/ep^2$ where $e$ is the fundamental unit of charge. The dwell times were varied typically in the range from 0.25 μs to 25 μs, with sub-ns precision (i.e. below 1% uncertainty). Imaging of the high-fluence implant squares with a 5 keV SEM beam (Fig. 1a) was used to quantify the effects of backlash and drift etc show that the actual area is sufficiently close to the nominal area that the dominant contribution to the uncertainty in $n$ is the precision on the current measurement. The sample was not annealed, i.e., no attempt was made to activate the Ga ions substitutionally and the collateral damage due to the implantation was not repaired. The sample implantation array was designed in order to provide regions of varying density and feature size, for tests of sensitivity and resolution, respectively. Each individual sample area contained rows of gallium implanted rectangles of varying width and density with spacing of 310 nm. All rectangles were $m$=100 FIB spots high (99$p$=1550nm). Each row began and ended with a 100x100 spot square = 1550x1550 nm² marker square with implantation density of $n$=80.1 ions nm$^{-2}$, to aid in location of the row region. We will concentrate discussion in this work on three rows in particular. The sensitivity test row (which we label sample S) was composed of rectangles of fixed width ($m$=100, $d$=1550nm) and varying density ranging from 40.1 to 0.06 nm$^{-2}$ (see Table 1). The resolution test row (sample R) had fixed density of 40 nm$^{-2}$ and rectangles of varying width from 50 down to a single line (with each rectangle repeated three times). Finally, the "lines" row (sample L) consisted of single lines ($m$=1) of FIB spots of decreasing density. In Table 1 the densities for sample L are specified such that $np^2$ is the number of ions per spot, but the actual peak density depends on the point spread function (PSF) for a single FIB spot. Although the second thinnest feature in sample R was specified as 2 lines, there was probably a rounding error in software that made this feature a single line.

A similar sample was prepared on a 30 nm thick single-crystal silicon membrane for use in energy dispersive X-ray spectroscopy (EDX) measurements in a Scanning Transmission Electron Microscope (STEM). The implanted regions consisted of 100x100 spot squares with $n$=8.3 to 1.2 ions nm² at a logarithmic step down in factors of 1.383. Other parameters were kept the same as the bulk implants such as $p$=15.64 nm, and no chromium marks were required.

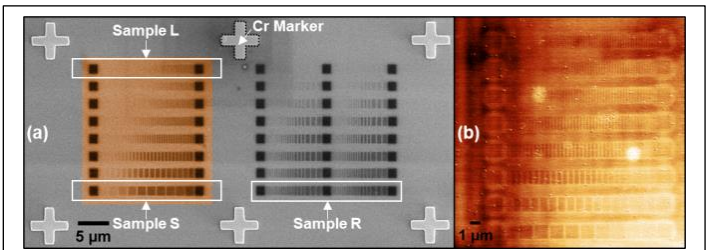

**Fig 1. Sample. a)** Scanning electron microscopy (SEM) image containing the regions of interest. Chromium marks can be seen as brighter crosses around the ion implanted regions. The greyscale gradient observed in the ion implanted geometries is a combination of minor height difference due to sputtered silicon atoms as well as changes in secondary electrons yield. Regions marked as sample "R", "S", and "L" correspond to patterns used for, respectively, XRF resolution, sensitivity, and limit of detection experiments described in the text. The overlaid orange-colored region corresponds to the Kelvin Probe Force Microscopy (KPFM) scanned area, which output in the potential channel signal with a drive amplitude of 7.5V can be seen in **b)**.

The sample was also imaged with Scanning Probe Microscopy (Kelvin Force - KPFM), Fig. 1b. The KPFM contrast mechanism is unclear since KPFM is sensitive to the work function and thus to subsurface charge, but the samples are unannealed and the dopants are unactivated. The images show clear contrast where the implants have occurred, and this could be due to a combination of the presence of dopants and the implantation damage. It is notable that there is a clear halo around the features in both the SEM and the KPFM, indicating that the FIB spot includes a component that is significantly larger than the nominal spot size, most often called ion beam tail.

*II.2. XRF Experiments*

Our nanoimaging experiments were conducted at the Hard X-ray Nanoprobe (HXN) beamline within NSLS-II, utilizing the Multi-Layer Laue (MLL) optics for nano-focusing. Detailed information on our instrumentation is available elsewhere[18–20]. In brief, HXN used an IUV20 undulator as a photon source and higher-order harmonics were rejected using an Rh-striped collimating mirror. A Si-111 double-crystal monochromator created a monochromatic beam with an energy of 13.5 keV (above the Ga K-edge), which was subsequently focused onto a secondary source aperture using both vertical and horizontal focusing mirrors. A nano-focused X-ray beam was generated by positioning MLLs vertically and horizontally. The resulting spot is approximately square, and fitting to knife-edge scans in each direction yields a full-width-at-half-maximum (FWHM) focus size of 15.3 × 16.9 nm², while using ptychography reconstruction[20] we obtain a FWHM size of 13.9 × 12.3 nm². An order sorting aperture (OSA) was placed between the sample and MLLs to block higher-order diffractions from the lenses. We performed raster scans of the sample stage to collect X-ray Fluorescence (XRF) and transmission data at each position, with an exposure time of typically 1 second. A three-channel silicon drift detector (Vortex, Hitachi) was positioned at a 90-degree angle relative to the sample to collect the XRF spectrum at each point. A transmission area detector (MERLIN, Quantum Detectors) was also placed at 0.5 m from the sample to capture far-field probe intensity. The integrated pump flux used was $5\times10^8$ photons s$^{-1}$.

**Table 1.** Sample design.

| Sample | Varying density, $n$ (ions nm$^{-2}$), or width, $m$ (number of lines) | Fixed $n$ or $m$ |
|---|---|---|
| S | $n$= 40.1, 21.0, 11.0, 5.76, 3.02, 1.58, 0.829, 0.434, 0.228, 0.0630 [log step down in factors of 1.9] | $m$= 100, $d$= 1550nm |
| R | $m$= 50, 25, 10, 5, 3, 2, 1 [$d$= ($m$-1)$p$] | $n$= 40.1 |
| L | $n$= 40.1, 35.9, 32.2, 28.8, 25.9, 23.2, 20.8, 18.6… 0.063 [log step down in factors of 1.116] | $m$= 1 |

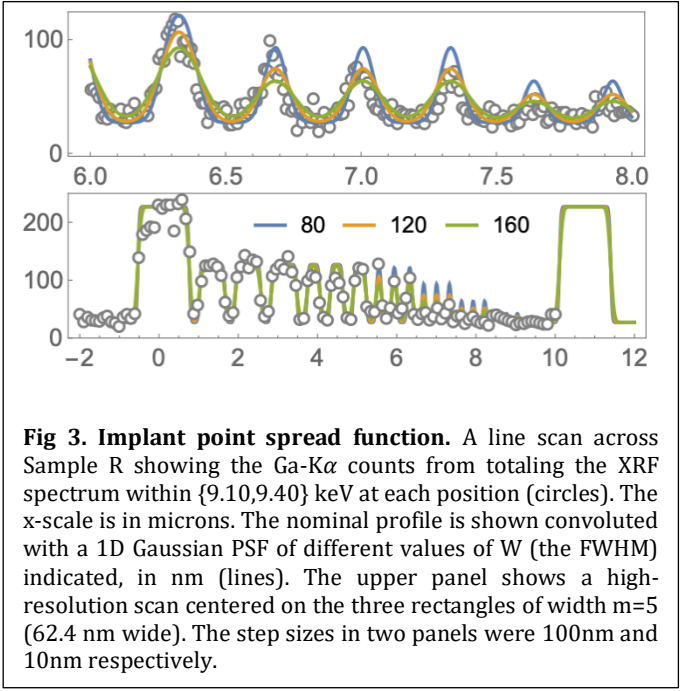

**Fig 3. Implant point spread function.** A line scan across Sample R showing the Ga-K$\alpha$ counts from totaling the XRF spectrum within {9.10,9.40} keV at each position (circles). The x-scale is in microns. The nominal profile is shown convoluted with a 1D Gaussian PSF of different values of W (the FWHM) indicated, in nm (lines). The upper panel shows a high-resolution scan centered on the three rectangles of width m=5 (62.4 nm wide). The step sizes in two panels were 100nm and 10nm respectively.

Before locating the region of interest, the sample was cleaned with acetone to remove any surface contaminants and cleaved to a manageable size. Filtering scans for Cr markers allowed to locate the arrays of interest after reshaping and refocusing of the X-ray beam. The Cr markers are expected to have very sharp edges due to the hard-mask lithography technique used, and these were used to get the sample into the focal plane of the beam. Scanning across a Cr marker line we obtained a 1D error function for the Cr XRF signal with minimum $W$ of 61 nm. Translating the sample along the beam increased $W$ by 5 nm per micron of translation.

*II.3. EDX Experiments*

EDX spectra were acquired (Bruker X-Flash dual-detectors) with electron-induced photons generated in a STEM (Thermo Scientific Talos F200i). The STEM was operated at 200 keV (quoted nominal imaging resolution equals to 136.5 pm) using a 70$\mu$m condenser lens aperture, resulting in an on-screen current of 0.386 nA. The EDX acquisition field of view was fixed at 1000x1000 nm$^2$, centered on the implanted squares, and the total integrated spectrum for an acquisition time of 300s was accumulated.

## III. RESULTS

First, we performed resolution tests using sample R. As shown in Fig. 2, $W$=120 nm which is somewhat higher than the value obtained from scanning over Cr markers above, and higher also than the nominal FIB implantation spot size (rms=15 nm convolved with rms straggle of 9 nm, produces an expected FWHM $W$=29nm - NB: for circular Gaussian spots of rms radius $w$, the edge of a 2D region produced with a pitch of $p<<w$ has a 1D error-function profile corresponding to a Gaussian with FWHM of $W = 2ln2w = 1.66w$). Evidently there is a significant tail in the beam profile around the FIB spot. It is unlikely that the X-ray beam spot is the cause of this large PSF since the focus was checked by scanning over an EBL-defined chromium marker about 5 μm away shortly beforehand and defocusing to a 120 nm spot would have required 24 μm translation along the beam.

We then investigated sample S for sensitivity tests as shown in Fig. 3. The Ga fluorescence from the implanted regions is obvious in the map of Ga fluorescence intensity as a function of position in the inset of Fig. 3. The mask regions chosen are at least 160 nm from the edge of the implanted area in order to avoid edge effects caused by the finite PSF mentioned above. The Ga spectral weight increases monotonically with the implant density, while the other background features (due to metal components in the focusing optics etc) remain constant. Sample S was imaged three times in succession (with slightly varying field of view in each image), and the results are shown in Fig. 4. The experiment is repeatable with the distributions plainly overlapping. The distribution of pixel values is very well separated from the background for a density of 40 nm$^{-2}$, meaning that if a single pixel contained an average of 40 impurities per nm$^2$ then it would be easily distinguishable from background. For 21 and 11 nm$^{-2}$ the distributions become decreasingly spaced from the background.

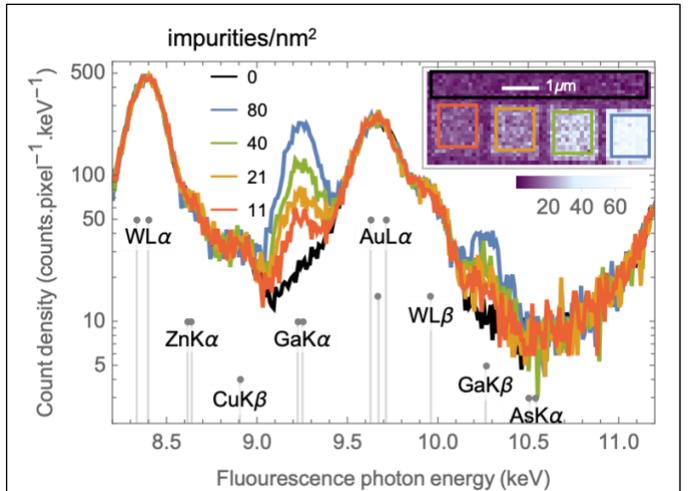

**Fig 2. XRF spectrum for different implant densities.** Sample S was imaged with a 1 s integration time per pixel, and at each pixel the acquired XRF spectrum was fitted with Gaussian peaks for the K and L transitions of the elements indicated (NB the heights of the gray labels for x-ray transitions have no meaning). The inset shows the integrated Ga-K$\alpha$ counts from the fits as a function of position. The main figure shows the mean spectra averaged over the 5 mask regions indicated on the inset, with Ga densities in impurities/nm$^2$ indicated by the legend. The integrated counts have been divided by the channel width to obtain a count density, so that the area under the Ga-K$\alpha$ peak in counts corresponds with the pixel values in the inset. No smoothing has been applied.

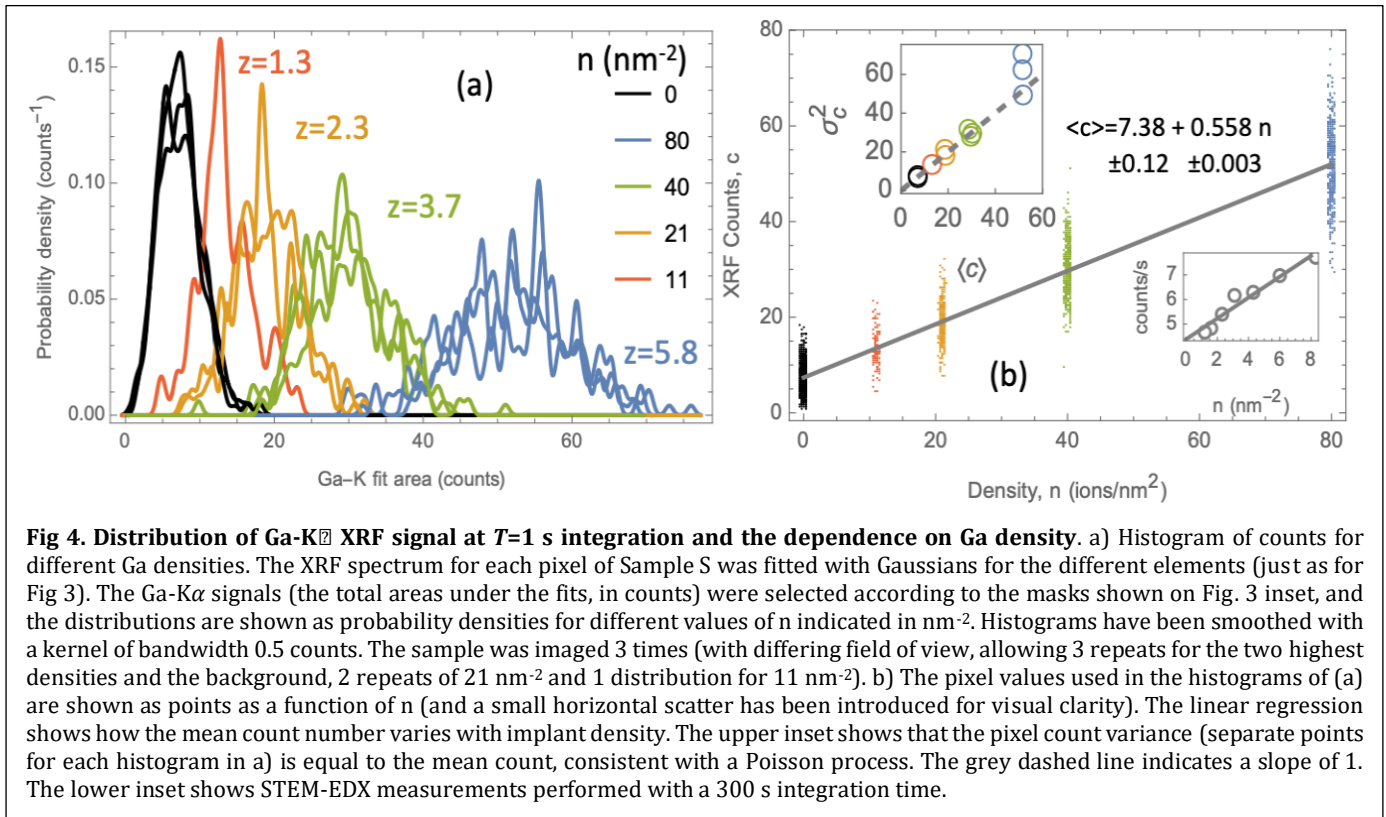

**Fig 4. Distribution of Ga-Kα XRF signal at $T=1$ s integration and the dependence on Ga density.** a) Histogram of counts for different Ga densities. The XRF spectrum for each pixel of Sample S was fitted with Gaussians for the different elements (just as for Fig 3). The Ga-K$\alpha$ signals (the total areas under the fits, in counts) were selected according to the masks shown on Fig. 3 inset, and the distributions are shown as probability densities for different values of n indicated in nm$^{-2}$. Histograms have been smoothed with a kernel of bandwidth 0.5 counts. The sample was imaged 3 times (with differing field of view, allowing 3 repeats for the two highest densities and the background, 2 repeats of 21 nm$^{-2}$ and 1 distribution for 11 nm$^{-2}$). b) The pixel values used in the histograms of (a) are shown as points as a function of n (and a small horizontal scatter has been introduced for visual clarity). The linear regression shows how the mean count number varies with implant density. The upper inset shows that the pixel count variance (separate points for each histogram in a) is equal to the mean count, consistent with a Poisson process. The grey dashed line indicates a slope of 1. The lower inset shows STEM-EDX measurements performed with a 300 s integration time.

When implantation features are extended and regularly repeated, they become easier to see. Sample L consists of single lines of FIB spots. As before, assuming that the FIB spot size is much larger than the pitch the implantation feature is a line with uniform density along the line and approximately Gaussian profile in the transverse direction of FWHM $W$. If the spots each contain $np^2$ impurities, the peak density is $n_0 = 4n\frac{p}{W}\sqrt{\ln 2/\pi}$. The values of $n$ in sample L are given in Table 1. Let us assume that the PSF of 120 nm from Fig. 2 is equal to $W$, which leads to peak densities of 9.8, 8.6, 7.5, 6.5, 5.7, 4.9, 4.3, 3.8 nm$^{-2}$, which, being below the background threshold determined above, are rather hard to see in the raw image (not shown). However, when the image is integrated in the direction parallel to the lines (16 pixels), the features become obvious as seen in Fig. 5. Effectively the acquisition time has been increased to 16 sec, and the fact that the features repeat regularly makes them clear. The weakest feature in Fig. 5 has 650 impurities in the X-ray spot at its peak.

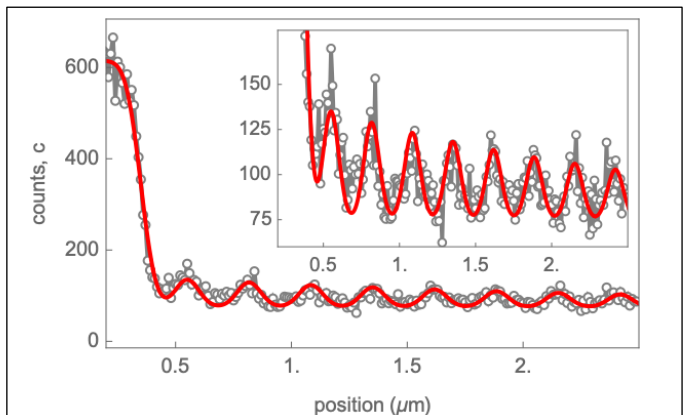

**Fig 5. Integrated signal from Sample L.** The sample consists of lines of constant spacing and decreasing implant density according to Table 1. A 2D image was taken and the end marker is clearly visible, but the lines are indistinct (not shown). Integrating the image over the vertical direction along the lines (with a small shear of 10º due to the slight misalignment of the sample to the scan axis), produces the data shown (the inset is a zoom). The red line is a convolution of the nominal density profile with a PSF of FWHM 120nm. The smallest density feature in Fig. 5 has peak density 3.8 nm$^{-2}$, corresponding to 650 Ga impurities for a 171 nm$^2$ spot.

## IV. DISCUSSION

The mean count rate detected for a uniform density, $n$, follows $\langle c(n) \rangle / T = b + an$ where $T$ is the integration time, $b$ is the mean background count rate and $a$ is the areal sensitivity. The best fit line on Fig. 3 for $T$=1 s gives $b$=7.38±0.12 s$^{-1}$ and $a$=0.558±0.003 nm$^2$ s$^{-1}$. Note that the inferred value of $a$ is not systematically affected by the X-ray beam size when the density of impurities is uniform - a defocussed spot produces a weaker pumping of a correspondingly larger number of impurities and ultimately the same number of fluorescence photons - and nor does the spot size affect the background. For comparison, in the EDX measurement of Fig. 4, $a$=0.42 nm$^2$ s$^{-1}$ and $b$=4.4 s$^{-1}$ which is strikingly similar.

The areal sensitivity, $a = \eta \sigma P$, is simply proportional to the quantum efficiency, $\eta$, the cross-section, $\sigma$, and the integrated pump flux in photons per unit time, $P$. For a single isolated cluster of $N$ impurities much smaller than the X-ray spot, $\langle c(N) \rangle / T = b + N/\tau$ where $\tau$ is the average time between fluorescence photons from a single impurity, and $1/\tau = \eta \sigma F_0$ where $F_0$ is the photon flux in the centre of the pump beam (where the cluster is), i.e. $\tau = A/a$ where $A$ is the X-ray spot size (defined so that $F_0 = P/A$). For the optimum XRF spot mentioned above, $A$ = 171 nm$^2$, then one Ga atom gives a count every $\tau$=306 s on average, which indicates the pixel integration time needed to detect a single impurity. In this experiment $b \gg 1/\tau$, meaning that for a single Ga impurity the background would swamp the signal.

Using the equation for $z$ above, and the fact that as shown in the inset in Fig. 4b the variance is equal to the count rate as expected for a Poisson point process, and solving for $N$ to obtain the detection limit,

$$N = \frac{z^2 \tau}{2T} \left( 1 + \sqrt{1 + 8bT/z^2} \right)$$

This formula has two regimes: the high background regime (large $b$) in which case $N = z\tau\sqrt{2b/T}$ so that the detection limit scales inversely with the root of the integration time; and the low background regime (small $b$) where $N = z^2 \tau / T$ which scales inversely with integration time. For $z$=2 and our experimental situation from Fig. 4, $T$=1 s $> z^2/8b = 0.068\ s$, we are in the high background regime. For $\tau$ =306 s and $T$=1 s we obtain a minimum detectable $N$ of 3000 impurities in a pixel ($N/A$=18 nm$^{-2}$). This number of impurities is equivalent to 350 zeptogrammes or a 4 nm cube of solid gallium.

Single atom detection is achieved with an integration time of $T = z^2 \tau (1 + 2b\tau)$, which can be decreased by decreasing $b$ or decreasing $\tau$, either by increasing the incident X-ray beam irradiance, $F_0$, or the collection efficiency through $\eta$. For example, the gold fluorescence seen in Fig. 3 (dominates $b$) might be reduced by a factor of 20 by choosing different materials in the X-ray optics. Instruments planned for new multiband-achromat storage ring light sources (that have either been recently constructed or are under construction) promise increased brilliance. Improvements in the focal spot flux are expected of the order of $10^{12}$ photons s$^{-1}$ in a 20 nm spot[21], which would lead to an improvement in $F_0$ by a factor of 800. Improving the flux density in the focus (and hence also $\tau$) by this much would then allow single atom detection $N$=1 with an integration time of just $T$=2 s (still background limited). Even this very high X-ray flux is not expected to produce radiation damage in bulk Si[22,23], but any damage that is produced will be far less than that caused by the implantation, which must anyway be healed and the impurities activated by annealing.

Although the requirements on $b$ and $\tau$ just described for imaging single atoms with unknown placement are stretching, the outlook for imaging regular patterns of impurities also often used in quantum technologies[24,25] is much more optimistic, since that allows better image processing techniques such as Bayesian inference to locate and analyse the pattern, and often structures with many atoms are used. Furthermore, patches of approximately $10^3$ impurities in a sheet of density approximately 1-2 nm$^{-2}$ have been used to construct quantum dots used in sensing of the charge/spin state of neighbouring isolated single impurity qubits[26], and it is just as important to characterize the dot as the qubit. This density is below the sensitivity limit determined from Fig. 4, but very close to the density of the weakest features observed in Fig. 5.

## V. CONCLUSION

XRF provides a sensitive and high-resolution capability for materials for quantum technology. Although other techniques are available with higher sensitivity (e.g., the EDX measurement of Fig. 4, or the KPFM result of Fig. 1 has visible features even well below 1 nm$^{-2}$, or X-ray induced STM[12], which senses single atoms on surfaces), XRF gives chemical specificity and sensitivity to buried features without destructive sample preparation. In addition to providing the means to locate dopants in qubits and other devices, X-ray characterization in the limit of very few, or even single, dopant atoms have the potential to provide structural and chemical insight that has not previously been possible. A particular challenge has been that the structure of defect complex structures, for example G-, T- W- and C-complexes in Si, which have been explained theoretically[2] but which have not been imaged directly, and which can have many related configurations of varying degrees of energetic favorability[27,28]: X-ray spectroscopy at the single-ion level has the potential to allow the structure predicted by calculations to be compared with experimental X-ray spectroscopy results. Ultimately X-ray studies could be conducted on samples or compared with single-ion optical or electrical studies. Extended X-ray absorption fine structure (EXAFS) measurements have, to date, required large ensembles of ions, with doping levels of $10^{20}$ cm$^{-3}$ or higher[29]. EXAFS requires a large range of incident energies and may thus be regarded as a long-term goal for the few-ions-scale nanoprobe. Studies drawing on near-edge spectroscopy require a narrower X-ray range of incident energies and would thus be more immediately available. Near edge studies can be used to probe the charge state of impurities, for example, but previously only at high dopant levels in MgO[30]. X-ray fluorescence therefore provides an important tool for chemical species-specific imaging at the nanoscale, and significant improvements are expected with near-future facility upgrades.

## Acknowledgment

MGM and SKC acknowledge financial support from Engineering and Physical Sciences Research Council (EPSRC) [Grant No. EP/X018989/1]. TG and VZ acknowledge support from EPSRC [DTP, Grant No. EP/T518050/1]. ATI-based authors would like to thank the UK National Ion Beam Centre [EPSRC, Grant No. EP/X015491/1] and UCL-based authors were financially supported by EPSRC [Grant No. EP/W000520/1, EP/R034540/1 and EP/V027700/1]. NK thanks the EPSRC Centre for Doctoral Training in Advanced Characterisation of Materials [Grant No. EP/S023259/1]. PGE acknowledges support from the U.S. Department of Energy Office of Basic Energy Sciences through contract DE-FG02-04ER46147. EDX analysis was supported by EPSRC [Grant No. EP/V036327/1].

## References

1. McCallum, J. C., Johnson, B. C. & Botzem, T. Donor-based qubits for quantum computing in silicon. *Appl Phys Rev* **8**, (2021).
2. Zhang, G., Cheng, Y., Chou, J.-P. & Gali, A. Material platforms for defect qubits and single-photon emitters. *Appl Phys Rev* **7**, (2020).
3. Groot-Berning, K. *et al.* Deterministic single-ion implantation of rare-earth ions for nanometer-resolution color-center generation. *Phys Rev Lett* **123**, 106802 (2019).
4. Cassidy, N. *et al.* Single ion implantation of bismuth. *physica status solidi (a)* **218**, 2000237 (2021).
5. Jakob, A. M. *et al.* Deterministic Shallow Dopant Implantation in Silicon with Detection Confidence Upper-Bound to 99.85% by Ion–Solid Interactions. *Advanced Materials* **34**, 2103235 (2022).
6. Räcke, P., Meijer, J. & Spemann, D. Image charge detection of ion bunches using a segmented, cryogenic detector. *J Appl Phys* **131**, (2022).
7. Mądzik, M. T. *et al.* Precision tomography of a three-qubit donor quantum processor in silicon. *Nature* **601**, 348–353 (2022).
8. Stock, T. J. Z. *et al.* Atomic-scale patterning of arsenic in silicon by scanning tunneling microscopy. *ACS Nano* **14**, 3316–3327 (2020).
9. Suenaga, K. *et al.* Element-selective single atom imaging. *Science (1979)* **290**, 2280–2282 (2000).
10. Voyles, P. M., Muller, D. A., Grazul, J. L., Citrin, P. H. & Gossmann, H.-J. Atomic-scale imaging of individual dopant atoms and clusters in highly n-type bulk Si. *Nature* **416**, 826–829 (2002).
11. Davies, T. E. *et al.* Experimental methods in chemical engineering: Scanning electron microscopy and X-ray ultra-microscopy—SEM and XuM. *Can J Chem Eng* **100**, 3145–3159 (2022).
12. Ajayi, T. M. *et al.* Characterization of just one atom using synchrotron X-rays. *Nature* **618**, 69–73 (2023).
13. D'Anna, N. *et al.* Non-Destructive X-Ray Imaging of Patterned Delta-Layer Devices in Silicon. *Adv Electron Mater* **9**, 2201212 (2023).
14. Laforce, B. *et al.* Assessment of ovarian cancer tumors treated with intraperitoneal cisplatin therapy by nanoscopic X-ray fluorescence imaging. *Sci Rep* **6**, 29999 (2016).
15. Shashkova, S. & Leake, M. C. Single-molecule fluorescence microscopy review: shedding new light on old problems. *Biosci Rep* **37**, BSR20170031 (2017).
16. Hashizume, T., Heike, S., Lutwyche, M. I., Watanabe, S. & Wada, Y. Atom structures on the Si (100) surface. *Surf Sci* **386**, 161–165 (1997).
17. Webb, R. P. & Wilson, I. H. An extension to the projected range algorithm (PRAL) to give energy deposition profiles'. in *Proc. 2nd Int. Conf. Simulation of semiconductor devices and processes (eds K. Board and DRJ OWen), Pineridge Press swansea UK* (1986).
18. Li, L. *et al.* X-Ray Nanoimaging: Instruments and Methods III, p. 103890U. *International Society for Optics and Photonics* (2017).
19. Nazaretski, E. *et al.* Design and performance of an X-ray scanning microscope at the Hard X-ray Nanoprobe beamline of NSLS-II. *J Synchrotron Radiat* **24**, 1113–1119 (2017).
20. Yan, H. *et al.* Multimodal hard x-ray imaging with resolution approaching 10 nm for studies in material science. *Nano Futures* **2**, 011001 (2018).
21. Maser, J. *et al.* Design of the in situ nanoprobe beamline at the Advanced Photon Source. in *X-Ray Nanoimaging: Instruments and Methods V* vol. 11839 15–21 (SPIE, 2021).

22. Polvino, S. M. *et al.* Synchrotron microbeam x-ray radiation damage in semiconductor layers. *Appl Phys Lett* **92**, (2008).
23. Zhang, J. *et al.* Study of X-ray radiation damage in silicon sensors. *Journal of Instrumentation* **6**, C11013 (2011).
24. Le, N. H., Fisher, A. J. & Ginossar, E. Extended Hubbard model for mesoscopic transport in donor arrays in silicon. *Phys Rev B* **96**, 245406 (2017).
25. Wang, X. *et al.* Experimental realization of an extended Fermi-Hubbard model using a 2D lattice of dopant-based quantum dots. *Nat Commun* **13**, 6824 (2022).
26. Broome, M. A. *et al.* High-fidelity single-shot singlet-triplet readout of precision-placed donors in silicon. *Phys Rev Lett* **119**, 046802 (2017).
27. Durand, A. *et al.* Broad diversity of near-infrared single-photon emitters in silicon. *Phys Rev Lett* **126**, 083602 (2021).
28. Baron, Y. *et al.* Detection of single W-centers in silicon. *ACS Photonics* **9**, 2337–2345 (2022).
29. Newman, B. K. *et al.* Extended X-ray absorption fine structure spectroscopy of selenium-hyperdoped silicon. *J Appl Phys* **114**, (2013).
30. Tanaka, I. *et al.* Identification of ultradilute dopants in ceramics. *Nat Mater* **2**, 541–545 (2003).